\documentclass{article}
\usepackage[a4paper, total={6in, 8in}]{geometry}

\usepackage[english]{babel}
\usepackage[utf8x]{inputenc}
\usepackage[T1]{fontenc}

\usepackage{amsmath}
\usepackage{siunitx}
\let\svqty\qty
\usepackage{physics}
\let\qty\svqty
\usepackage{bm}
\usepackage{graphicx}
\usepackage{float}
\usepackage{subcaption}
\usepackage[colorinlistoftodos]{todonotes}
\usepackage[colorlinks=true, allcolors=black]{hyperref}
\usepackage{listings}
\usepackage[affil-it]{authblk}
\usepackage{tikz}
\usepackage{overpic}
\usepackage{cleveref}
\usepackage{parskip}
\usepackage{svg}

\title{Antiferromagnetic half-skyrmions electrically generated and controlled at room temperature}
\author[1]{\small O. J. Amin}
\author[1]{\small S. F. Poole}
\author[1,2,3]{\small S. Reimers}
\author[1]{\small L. X. Barton}
\author[2]{\small F. Maccherozzi}
\author[2]{\small S. S. Dhesi}
\author[4]{\small V. Nov\'ak}
\author[4]{\small F. K\v r\'i\v zek}
\author[1]{\small J. S. Chauhan}
\author[1]{\small R. P. Campion}
\author[1]{\small A. W. Rushforth}
\author[1,4]{\small T. Jungwirth}
\author[5]{\small O. A. Tretiakov}
\author[1]{\small K. W. Edmonds}
\author[1]{\small P. Wadley}
 
\affil[1]{\footnotesize School of Physics and Astronomy, University of Nottingham, Nottingham, NG7 2RD, United Kingdom}
\affil[2]{\footnotesize Diamond Light Source, Chilton, OX11 0DE, United Kingdom}
\affil[3]{\footnotesize Johannes Gutenberg Universit\"at Mainz, Institut f\"ur Physik,
Staudingerweg 7, 55128 Mainz, Germany}
\affil[4]{\footnotesize Institute of Physics, Czech Academy of Sciences, 162 00 Praha 6, Prague, Czech Republic}
\affil[5]{\footnotesize School of Physics, The University of New South Wales, Sydney 2052, Australia}

\begin{document}

\maketitle

\begin{abstract}
\textbf{Topologically protected magnetic textures, such as skyrmions, half-skyrmions (merons) and their antiparticles, constitute tiny whirls in the magnetic order. They are promising candidates for information carriers in next-generation memory devices, as they can be efficiently propelled at very high velocities using current-induced spin torques \cite{Tomasello2014,Geng2017,Gobel2019,Gobel2021,Yu2021,Juge2021b}. Antiferromagnets have been shown to host versions of these textures, which have gained significant attention because of their potential for terahertz dynamics, deflection free motion, and improved size scaling due to the absence of stray field \cite{Jani2021,Barker2016,Sort2006,Wu2011,Kolesnikov2018,Chmiel2018}. 
Here we show that topological spin textures, merons and antimerons, can be generated at room temperature and reversibly moved using electrical pulses in thin film CuMnAs, a semimetallic antiferromagnet that is a testbed system for spintronic applications \cite{Wadley2016,Grzybowski2017,Olejnik2017,Olejnik2018,Wadley2018,Kaspar2021,Zubac2021}. The electrical generation and manipulation of antiferromagnetic merons is a crucial step towards realizing the full potential of antiferromagnetic thin films as active components in high density, high speed magnetic memory devices.}
\end{abstract}

The defining feature of a topological texture is a non-zero winding number, also called topological charge. In magnetic systems, this is a measure of the integer number of times the order parameter wraps around a unit sphere (Bloch sphere) when integrated over the texture volume. Magnetic textures with different winding numbers are topologically distinct and cannot be easily changed from one topological state to another. This provides strong protection against perturbation and prevents their collapse even at ultrasmall sizes \cite{Kosterlitz1973}. In ferromagnetic (FM) thin films, where the magnetization defines the topological charge, $Q$, skyrmions ($Q=\pm1$) and merons ($Q=\pm1/2$) have been generated and controlled using external fields and current-induced spin torques \cite{Shinjo2000,Klaui2006,Jiang2015,Coisson2016,Gao2019,Gao2020b,Li2021b,Wang2021}. However, their implementation into practical devices has been limited by the presence of gyrotropic forces, originating from their topology, that cause an unwanted transverse deflection to their current-driven motion \cite{Zhang2020}.
\par

\begin{figure}[ht]
    \centering
    \includegraphics{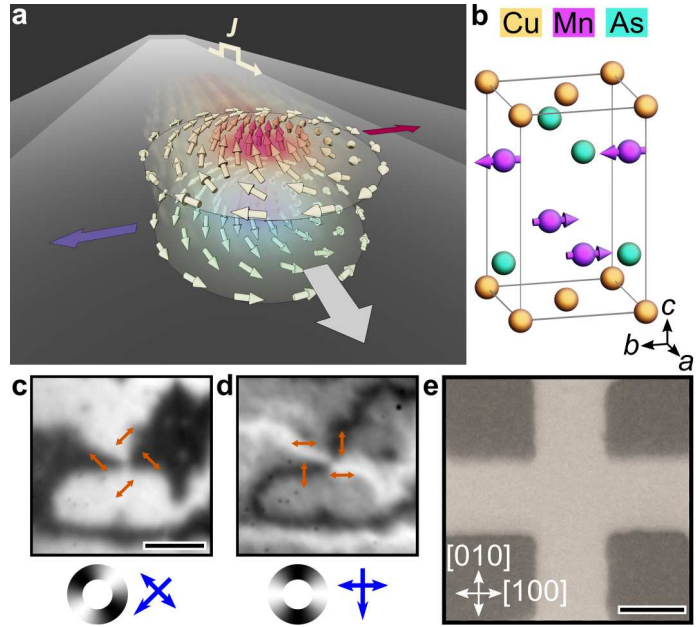}
    \caption{AF textures in CuMnAs. a, Spin structure and force acting on an  AF Bloch-type meron under an applied current pulse $\bm{J}$. b, Unit cell and magnetic structure of CuMnAs. c, d, XMLD-PEEM images of a vortex structure in CuMnAs. The blue single- and double-headed arrows indicate the x-ray incidence and polarization vectors, while the colour wheels and red double-headed arrows indicate the spin axis orientation inferred from the XMLD contrast. The scale bar corresponds to \SI{1}{\micro\meter}. e, Optical image of the device structure used for electrical pulsing. The spatial scale bar corresponds to \SI{10}{\micro\meter}.}
    \label{fig1}
\end{figure}

This effect is alleviated in antiferromagnets (AFs) due to their compensated FM sublattices. The AF order parameter, called the N\'eel vector, is given by $\bm{L}=\bm{M}_{1}-\bm{M}_{2}$ in collinear AFs, where $\bm{M}_{1}$ and $\bm{M}_{2}$ are antiparallel sublattice magnetizations. Analogous to the magnetization in FMs, the N\'eel vector defines the N\'eel topological charge, $Q_{\text{N}}$, of the AF spin texture. However, current-induced gyrotropic forces, $F^{(k)}_{\text{gyro}}=G^{(k)}\hat{z}\times\bm{J}$, depend on the gyrocoupling constant $G=4\pi Q^{(k)}$, which is a function of the sublattice topological charge, $Q^{(k)}$. 
When integrated over the texture volume the gyrotropic forces fully compensate, resulting in a remaining generalized drag force, $F_{\text{drag}}=\Gamma\beta\bm{J}$, in the direction of the current, where $\beta$ is the sum of the current-induced spin torques acting on the AF topological texture \cite{Barker2016}.
The dynamics of the AF topological texture under the action of current-induced spin torques is  described by the generalized Thiele’s equation \cite{Thiele73, Tretiakov2008, Tveten2013},
\begin{equation}\label{Thiele_eqn}
    F^{i}=\mathcal{M}^{ij}\ddot{b}_{j}+ \alpha\Gamma^{ij} \dot{b}_{j},
\end{equation}
where the $b_j$ are the collective coordinates of the topological spin structure, $\mathcal{M}^{ij}$ is the mass tensor, and $\alpha\Gamma^{ij}$ characterizes the viscous friction.
The AF topological texture is propelled by this force at terminal velocity $v_{\parallel}=\beta\bm{J}$, parallel to the current direction, as shown schematically in Fig.~\ref{fig1}a.
\vspace{5pt}
\par
The incentive for finding host materials has led to the experimental observation of AF topological textures in complex systems, including synthetic AF heterostructures \cite{Kolesnikov2018} and FM/AF bilayers \cite{Sort2006,Wu2011,Chmiel2018}. Recently, the nucleation of merons and antimerons has been demonstrated in a pure AF material, $\alpha$-Fe$_2$O$_3$ \cite{Jani2021}. However, nucleation methods requiring either external magnetic fields or thermal cycling are impractical for integrable devices. It has also yet to be shown that, once generated, topological AF textures can be controllably moved. Here, we bridge this gap by showing that AF merons and antimerons can be electrically generated and controlled at room temperature in a conducting AF material, CuMnAs. CuMnAs has the required crystal symmetries to host current-induced N\'eel spin orbit torques (NSOTs), which can efficiently manipulate the AF order \cite{Wadley2016}. This makes it an ideal candidate material for AF-based spintronic devices.
\par
Our experiments are performed on tetragonal phase CuMnAs (space group P4/nmm), epitaxially grown on a lattice-matched GaP(001) substrate \cite{Krizek2020}. Below the N\'eel temperature of \SI{480}{\kelvin}, the Mn atoms form two magnetic sublattices which are stacked vertically in the crystallographic $c$-axis (Fig.~\ref{fig1}b). The magnetic easy axis lies in the \textit{ab}-plane due to magnetocrystalline anisotropy. The precise behaviour of the magnetic anisotropy is strongly influenced by the interface with the substrate and, depending on film thickness, can be tuned between in-plane uniaxial and biaxial anisotropies \cite{Janda2020}. 
\par
To resolve the magnetic structure we used X-ray magnetic linear dichroism (XMLD) combined with photoemission electron microscopy (PEEM). Figs.~\ref{fig1}c, d show an image of an AF vortex structure in CuMnAs. With the incident x-ray polarization along one of the biaxial easy axes (Fig.~\ref{fig1}c), the \SI{0}{\degree} and \SI{90}{\degree} AF domains appear as dark and light regions, respectively. When the polarization is at \SI{45}{\degree} to the easy axes, the contrast is largest at the AF domain walls (AFDWs), which appear either dark or light depending on the orientation of the spin at the domain wall centre (Fig.~\ref{fig1}d). The AF vortex is identified by the dark-light-dark-light pattern of the AFDWs along a circular path. While it is not possible to distinguish the sign of $\bm{L}$ by XMLD, the chirality of the vortex structure depends only on the spin axis rotation, determinable from the two x-ray configurations.
\par
 
\subsection*{Generation of merons and antimerons using electrical pulses}
\begin{figure}[ht]
    \centering
    \includegraphics{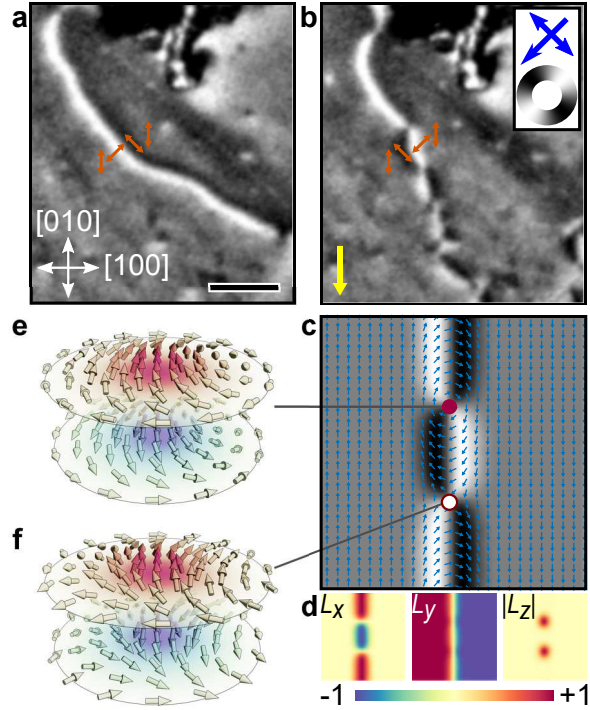}
    \vspace{-5pt}
    \caption{Generation of AF meron-antimeron pairs using an electrical pulse. a, XMLD-PEEM image of a \SI{180}{\degree} AFDW located in the top arm of the device. White and black contrast corresponds to N\'eel vector orientation perpendicular and parallel to the X-ray linear polarization (blue double-headed arrow in top right), respectively. The spin axis variation across the AFDW width is depicted by red arrows. The spatial scale bar corresponds to \SI{600}{\nano\meter}. b, The AFDW after applying a \SI{1}{\milli\second} electrical pulse with \SI{21}{\volt} (\SI{1.2e7}{\ampere\per\centi\meter\squared}) amplitude along the [0$\bar{1}$0] CuMnAs crystal direction (yellow arrow), showing sections of reversed chirality. The chirality changes are associated with AF vortices and antivortices. c, d Simulated XMLD-PEEM image and N\'eel vector heatmaps for an  AFDW showing chirality reversal. e, f, Characteristic Bloch-type meron (e) and antimeron (f) with out-of-plane core spin component, located at positions highlighted by red and white filled circles in (c).}
    \label{fig2}
\end{figure}

To investigate the current-induced generation and manipulation of topological spin structures, we fabricated a 4-arm cross device with \SI{10}{\micro\meter} arm width from a \SI{50}{\nano\meter} layer of CuMnAs (Fig.~\ref{fig1}e). In the patterned sample, the magnetic anisotropy is uniaxial, with domains oriented along the [010] crystalline axis. Fig.~\ref{fig2}a shows an XMLD-PEEM image of a \SI{180}{\degree} AFDW identified in one of the arms of the device. Incident x-rays have linear polarization vector along the [1$\bar{1}$0] CuMnAs crystal axis. The variation in $\bm{L}$ across the AFDW width is resolved as white and black contrast, corresponding to in-plane magnetic moments which are aligned perpendicular and parallel to the x-ray polarization, respectively, as indicated by the colour wheel in the top-right corner of Fig.~\ref{fig2}b.
\par
A \SI{1}{\milli\second} electrical pulse, with amplitude \SI{21}{\volt} (corresponding to a current density of \SI{1.2e7}{\ampere\per\centi\meter\squared}) was applied along the [010] CuMnAs axis of the device. Fig.~\ref{fig2}b shows the micromagnetic structure of the AFDW after the pulse was applied. Sections of the AFDW have the order of contrast reversed from white/black to black/white, corresponding to a reversal of chirality.
Monte-Carlo methods with magnetic anisotropy parameters typical for thin film CuMnAs \cite{Maca2017,Wang2021} were used to simulate the structure of the AFDW. The simulated dichroism image with overlaid top sublattice magnetization is shown in Fig.~\ref{fig2}c. Topological spin textures are seen at the AFDW chirality reversal points, with the positions of their cores highlighted by filled circles. At the position of the red filled circle, the spin texture has the characteristic structure of a Bloch-type meron - spins along its diameter form a Bloch-type \SI{180}{\degree} domain wall. Whereas the spin texture at the position of the white filled circle is an antimeron. There are two directions along which the diameter forms Bloch-type \SI{180}{\degree} domain walls, and two which form N\'eel-type \SI{180}{\degree} domain walls. The $L_{x}$, $L_{y}$ and $|L_{z}|$ N\'eel vector components are plotted as heatmaps in Fig.~\ref{fig2}d. Most notably, $|L_{z}|$ shows an out-of-plane spin orientation at the core of the vortex - as is required for topologically non-trivial textures. The full 3-dimensional spin structure of the AF meron and antimeron are shown in Figs.~\ref{fig2}e, f.
\vspace{5pt}

\subsection*{Electrical control of generated meron-antimeron pairs}
\begin{figure}[ht]
    \centering
    \includegraphics{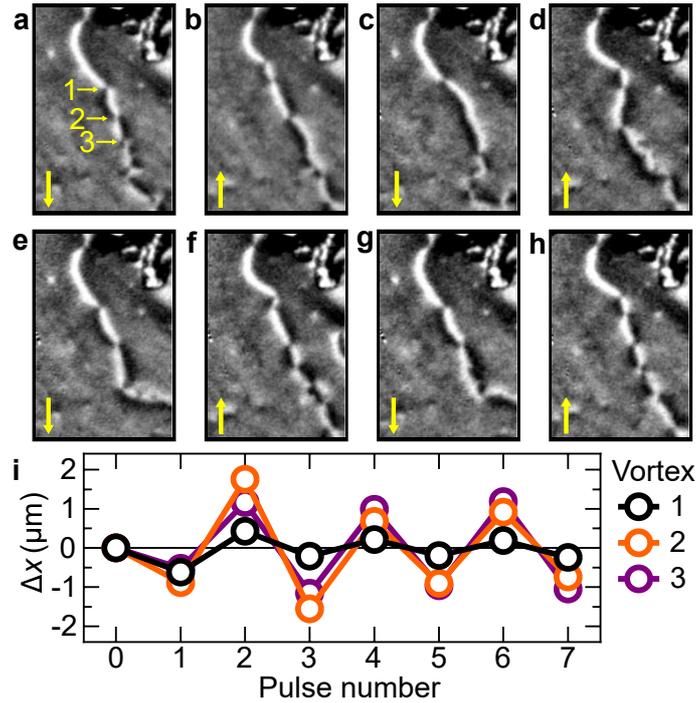}
    \caption{Movement of meron-antimeron pairs using electrical pulses. a-h, XMLD-PEEM images of the \SI{180}{\degree} AFDW after applying successive \SI{1}{\milli\second} \SI{21}{\volt} electrical pulses in the directions indicated by the yellow arrows. i, The average displacement of vortices 1 to 3 indicated in a, measured after each pulse.}
    \label{fig3}
\end{figure}

Subsequent \SI{1}{\milli\second} electrical pulses, with \SI{21}{\volt} amplitude, were applied in a sequence of alternating polarity along the [010] CuMnAs axis. Fig.~\ref{fig3} shows XMLD-PEEM images of the \SI{180}{\degree} AFDW after each pulse in the sequence. Yellow arrows show the current direction of the applied pulse. The AFDW state shown in Fig.~\ref{fig3}a corresponds to the same state shown in Fig.~\ref{fig2}b. After successive pulses of alternating polarity, shown in Figs.~\ref{fig3}b-h, the points of AFDW chirality reversal, where merons and antimerons are localized, are seen to move collectively in the direction of the applied pulse. For each image, the same meron-antimeron pair can be identified (see Supplementary Information) as the white/black, black/white (and vice versa) contrast reversal, indicating a change in the AFDW chirality, which necessitates the type of topological vortex at that point. The displacement, $\Delta x$, of three merons indicated in Fig.~\ref{fig3}a, along the [0$\bar{1}$0] direction, collinear to the pulse direction, is shown in Fig.~\ref{fig3}i for the sequence of 8 alternating polarity pulses. Their movement follows a reversible, repeatable pattern dependent on the pulse polarity. Well defined displacements between successive pulses are indicative of pinning sites, created by crystal defects and local strains, which determine the stable positions for the meron-antimeron pairs. We preclude thermal effects from causing the reversible motion, as these are an even function of the pulse polarity.
\vspace{5pt}

\subsection*{Observation of isolated meron-antimeron pairs on AFDW loops}
\begin{figure}[ht]
    \centering
    \includegraphics{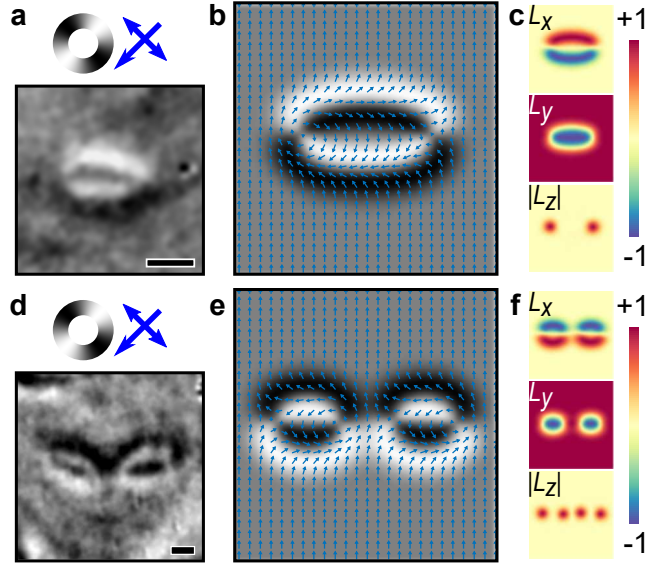}
    \caption{Isolated meron-antimeron pairs localized at the points of chirality reversal on \SI{180}{\degree} AFDW loops. XMLD-PEEM images of (a) single \SI{180}{\degree} AFDW loop and (d) double \SI{180}{\degree} AFDW loops. Spatial scale bars correspond to \SI{350}{\nano\meter}. Simulated XMLD-PEEM image (b, e) and N\'eel vector heatmaps (c, f) for both structures showing meron-antimeron pairs situated at points of AFDW chirality reversal.}
    \label{fig4}
\end{figure}

As well as the electrical generation and control of meron-antimeron pairs along a \SI{180}{\degree} AFDW, we observed, in distinct regions of the device, the formation of isolated pairs on \SI{180}{\degree} AFDW loops. Fig.~\ref{fig4}a, d show two XMLD-PEEM images of example \SI{180}{\degree} AFDW loops. On each loop there exist two points of chirality reversal corresponding to a meron and antimeron pair. This is revealed in the micromagnetic simulations of such structures shown in Fig.~\ref{fig4}b, e. In the case of two adjacent AFDW loops, shown in Fig.~\ref{fig4}d, a meron and antimeron in close proximity form a bimeron, depending on the polarity of each structure.
\par
\subsection*{Discussion}
For nucleation of meron-antimeron pairs to occur, the pulse amplitude must exceed a nucleation energy threshold and the energy must be supplied within short enough time so as not to be dissipated due to damping in the system. The magnetic dynamics in the system can be described by the following Landau-Lifshitz-Gilbert (LLG) equation:
\begin{equation}
\label{LLG}
\dot{\bm{m}}^{(k)}= \gamma \bm{m}^{(k)}\times\pdv{F}{\bm{m}^{(k)}} +\alpha \bm{m}^{(k)} \times \dot{\bm{m}}^{(k)} +\bm{\tau},    
\end{equation}
where $\bm{m}$ is the magnetic moment on sublattice $k$, $F$ is the magnetic functional of the system, $\alpha$ is the Gilbert damping constant, and $\tau$ corresponds to all of the current-induced spin torques. The Gilbert damping defines the time scale on which to supply the critical energy needed to create the meron-antimeron pair from topological vacuum. In the exchange-dominated limit, this critical energy is given by a topological invariant $8\pi Q_{\text{N}} A t$, where $A$ is the AF exchange constant and $t$ is the thickness of the sample~\cite{Tretiakov2007}. In our system this energy is typically higher as it involves all other interactions present.  We note that the meron-antimeron pairs are localized on the 180$^\circ$ AFDW, see Fig.~\ref{fig2}b, as this is where the nucleation energy is minimum.
\par
In AFs both meron and antimeron move in the direction of the current, in this particular case confined along the length of a \SI{180}{\degree} AFDW, as it again minimizes the energy in the system. This is attributed to the fact that in AFs gyrotropic forces completely cancel out from the two AF sublattices \cite{Barker2016}. The resulting force in Thiele's equation (Eqn.~\ref{Thiele_eqn}) is always in the current direction and is independent of the topological charge, i.e., it is the same for both meron and antimeron. Any type of spin torque, including current-induced NSOT originating from the crystal symmetry of CuMnAs, as well as spin transfer torques due to the gradient in $\bm{L}$, will contribute to the force. The pinning sites in the CuMnAs play an important role in determining the positions between which the merons and antimerons move, see Fig.~\ref{fig3}.
\par
The results presented demonstrate CuMnAs as a suitable electrically conducting AF host material for topological spin textures. Efficient generation and control of AF merons and antimerons using electrical pulses paves the way towards realizing practical AF-based solitonic devices. The role of magnetic anisotropy on stable meron size can be explored by tuning the CuMnAs epilayer thickness \cite{Janda2020} and the controlled fabrication of structural pinning sites \cite{Reimers2022} may be used to selectively determine distances between which merons and antimerons move. This could lead to the development of novel device architectures, combining constricted channels to nucleate isolated meron-antimeron pairs \cite{Jiang2015} that may be used to carry information on racetrack memories \cite{Tomasello2014}.

\section*{Methods}
\subsection*{Sample fabrication}
The \SI{50}{\nano \meter} CuMnAs film used for this study was grown by molecular beam epitaxy on a GaP buffer layer on a GaP(001) substrate. The film was capped with a \SI{3}{\nano \meter} Al film to prevent surface oxidation. The device structure shown in Fig.~\ref{fig1}e, with device arms aligned along the [100] and [010] CuMnAs crystal axes, was fabricated using optical lithography followed by ion etching.

\subsection*{PEEM imaging and electrical pulsing}
The photoemission electron microscopy measurements were performed on beamline I06 at Diamond Light Source. The X-ray beam was incident at a grazing angle of \SI{16}{\degree}, with the X-ray linear polarization in the plane of the film. The asymmetry, $\Delta=(I_{E_{1}}-I_{E_{2}})/(I_{E_{1}}+I_{E_{2}})$, between images obtained at the Mn $L_3$ absorption peak ($E_1$) and at \SI{0.9}{eV} below the $L_{3}$ peak ($E_2$), provides a map of the local spin axis with \SI{30}{\nano\meter} spatial resolution. The sample was mounted on a cartridge with four wire-bonded electrical contacts allowing electrical pulses to be applied in-situ within the PEEM vacuum chamber. After each pulse, XMLD-PEEM images of device regions were taken to map changes to the micromagnetic structures. The time between applying a pulse and acquiring an image exceeded \SI{120}{\second}, thus only non-transient changes to the magnetic state were observed.

\subsection*{Micromagnetic simulations}
We use Monte Carlo methods on a $51\times 51\times 2$ system of spins to simulate the magnetic textures observed in CuMnAs. The system has periodic boundary conditions in the first ($x$) and third ($z$) dimensions, and closed boundaries in the second ($y$). Each layer of the system (in $z$) corresponds to a magnetic sublattice, with AF coupling between the layers. The total energy of the system is defined as
 the sum of the exchange energy, the out-of-plane magnetic anisotropy, and the in-plane magnetic anisotropy. Spins have full rotational degrees of freedom with out-of-plane polar angle, $\theta$, and in-plane azimuthal angle, $\phi$. The system is initiated with in-plane textures and allowed to settle for $N\times$1e6 iterations.

\bibliography{main.bib}
\bibliographystyle{unsrt}

\section*{Acknowledgements}
We thank Diamond Light Source for the allocation of beamtime on beamline \texttt{I06} under Proposal nos. SI26255-1 and MM27845-1. The work was supported by the EU FET Open RIA Grant no 766566 and the UK Engineering and Physical Sciences Research Council [grant number EP/V031201]. PW acknowledges support from the Royal Society through a University Research Fellowship.

\section*{Author contributions}
PW, KWE and TJ conceived and led the project. OJA, LXB, VN, FK, JSC, RPC and AWR contributed to fabrication of materials and devices. OJA, SFP, SR, LXB, FM and SSD performed the XPEEM experiments and data analysis. OJA performed the micromagnetic simulations. OJA, KWE, OAT and PW wrote the manuscript with feedback from all authors.

\section*{Competing Interests}
The authors declare no competing interests.

\end{document}